\documentclass[%
 reprint,
 notitlepage,
superscriptaddress,
preprintnumbers,
nofootinbib,
 amsmath,amssymb,
 aps,
twocolumn, 
]{revtex4-2}
\usepackage{tikz}
\usepackage{verbatim}
\usepackage{graphicx}
\usepackage{dcolumn}
\usepackage{bm}
\usepackage{booktabs}

\usepackage{xcolor}
\usepackage{tabularx}
\usepackage{booktabs}
\usepackage{array}
\usepackage[colorlinks]{hyperref}
\hypersetup{
    colorlinks,
    linkcolor={red!50!black},
    citecolor={blue!50!black},
    urlcolor={blue!80!black}
}
\usepackage[normalem]{ulem}
\usepackage{physics}
\usepackage{comment}
\usepackage{multirow}
\usepackage{subfigure}
\usepackage{xspace}
\usepackage{slashed}

\usepackage{enumitem}

\newcommand{\La}{{\rm \Lambda}}
\newcommand\ddcalc{\textsf{DDCalc}\xspace}
\newcommand\directdm{\textsf{DirectDM}\xspace}
\newcommand\darkbit{\textsf{DarkBit}\xspace}
\newcommand\ddcalcthree{\textsf{DDCalc 3.0}\xspace}

\newcommand{\gambit}{\textsf{GAMBIT}\xspace}
\newcommand{\gambitlight}{\textsf{GAMBIT-light}\xspace}

\newcommand{\diver}{\textsf{Diver}\xspace}

\newcommand{\keVee}{{\rm \,keV_{\rm ee}}}

\newcommand{\Q}[2]{
  \if\relax\detokenize{#2}\relax
    \mathcal{Q}_{#1}
  \else
    \mathcal{Q}_{#1}^{(#2)}
  \fi
}

\newcommand{\C}[2]{
  \if\relax\detokenize{#2}\relax
    \mathcal{C}_{#1}
  \else
    \mathcal{C}_{#1}^{(#2)}
  \fi
}

\newcolumntype{Y}{>{\centering\arraybackslash}X}
\newcolumntype{Z}[1]{>{\centering\arraybackslash}m{#1}}
\usepackage{cellspace}
\graphicspath{{figures/}}

\begin{document}
\def\ppnp{Prog.\ Part.\ Nuc.\ Phys.}
\def\pdu{Phys.\ Dark Univ.}
\def\astropacific{Astron.\ Soc.\ Pacific Conf.\ Ser.}
\def\lnp{Lec.\ Notes in Physics}
\def\cpc{Comp.\ Phys.\ Comm.}
\def\jpg{J. Phys. G}
\def\ijmpa{Int.\ J.\ Mod.\ Phys.\ A}
\def\ijmpd{Int.\ J.\ Mod.\ Phys.\ D}
\def\epjc{Eur.\ Phys.\ J.\ C}
\def\nima{Nuc.\ Inst.\ Methods A}
\def\nimb{Nuc.\ Inst.\ Methods B}
\def\njp{New J.\ Phys.}
\def\rmp{Rev.\ Mod.\ Phys.}
\def\app{Astropart.\ Phys.}
\def\aj{AJ}%
\def\actaa{Acta Astron.}%
\def\araa{ARA\&A}%
\def\arnps{Ann.~Rev.~Nucl.~\& Part.~Sci.}%
\def\apj{ApJ}%
\def\apjl{ApJ}%
\def\apjs{ApJS}%
\def\ao{Appl.\ Opt.}%
\def\apss{Ap\&SS}%
\def\aap{A\&A}%
\def\aapr{A\&A~Rev.}%
\def\aaps{A\&AS}%
\def\azh{AZh}%
\def\pos{PoS}%
\def\baas{BAAS}%
\def\bac{Bull.\ Astr.\ Inst.\ Czechosl.}%
\def\caa{Chinese Astron.\ Astrophys.}%
\def\cjaa{Chinese J.\ Astron.\ Astrophys.}%
\def\icarus{Icarus}%
\def\jhep{JHEP}%
\def\jcap{JCAP}%
\def\jpsj{J.\ Phys.\ Soc.\ Japan}%
\def\jrasc{JRASC}%
\def\canjphys{Can.~J.~Phys.}
\def\apphys{Astropart.~Phys.}
\def\mnras{MNRAS}%
\def\memras{MmRAS}%
\def\na{New A}%
\def\nar{New A Rev.}%
\def\pasa{PASA}%
\def\pra{Phys.\ Rev.\ A}%
\def\prb{Phys.\ Rev.\ B}%
\def\prc{Phys.\ Rev.\ C}%
\def\prd{Phys.\ Rev.\ D}%
\def\pre{Phys.\ Rev.\ E}%
\def\prl{Phys.\ Rev.\ Lett.}%
\def\pasp{PASP}%
\def\pasj{PASJ}%
\def\qjras{QJRAS}%
\def\rmxaa{Rev. Mexicana Astron. Astrofis.}%
\def\skytel{S\&T}%
\def\solphys{Sol.\ Phys.}%
\def\sovast{Soviet~Ast.}%
\def\ssr{Space~Sci.\ Rev.}%
\def\zap{ZAp}%
\def\nat{Nature}%
\def\science{Science}%
\def\sci{\science}%
\def\iaucirc{IAU~Circ.}%
\def\aplett{Astrophys.\ Lett.}%
\def\apspr{Astrophys.\ Space~Phys.\ Res.}%
\def\bain{Bull.\ Astron.\ Inst.\ Netherlands}%
\def\fcp{Fund.\ Cosmic~Phys.}%
\def\gca{Geochim.\ Cosmochim.\ Acta}%
\def\grl{Geophys.\ Res.\ Lett.}%
\def\jcp{J.\ Chem.\ Phys.}%
\def\jgr{J.\ Geophys.\ Res.}%
\def\jqsrt{J.\ Quant.\ Spec.\ Radiat.\ Transf.}%
\def\memsai{Mem.\ Soc.\ Astron.\ Italiana}%
\def\nphysa{Nucl.\ Phys.\ A}%
\def\nphysb{Nucl.\ Phys.\ B}%
\def\physrep{Phys.\ Rep.}%
\def\physscr{Phys.\ Scr}%
\def\planss{Planet.\ Space~Sci.}%
\def\procspie{Proc.\ SPIE}%
\def\repprogphys{Rep.\ Prog.\ Phys.}%
\def\jpcrd{J. Phys. Chem. Ref. Data}%
\def\jphysb{J. Phys. B}%
\def\jphysd{J. Phys. D}%
\def\jphysconfseries{J. Phys. Conf. Series}%
\def\physrev{\pr}
\def\pr{Phys. Rev.}%
\def\josa{J. Opt. Soc. Amer. (1917-1983)}%
\def\josab{J. Opt. Soc. Amer. B}%
\def\pla{Phys. Lett. A}%
\def\plb{Phys. Lett. B}%
\def\os{Opt. Spectrosc. (Russ.)}%
\def\jas{J. Appl. Spectrosc.}%
\def\annp{Ann. Phys.}%
\def\sa{Spectrochim. Acta}%
\def\prsoca{Proc. R. Soc. London Ser. A}%
\def\zphysa{Z. Phys. A}%
\def\zphysb{Z. Phys. B}%
\def\zphysc{Z. Phys. C}%
\def\zphysd{Z. Phys. D}%
\def\zphyse{Z. Phys. E}%
\def\zphys{Z. Phys.}%
\def\adndt{Atom. Data Nuc. Data Tables}%
\def\jmolspec{J. Mol. Spectrosc.}%
\def\aphysb{Appl. Phys. B}%
\def\nim{Nuc. Inst. Meth.}%
\def\jphysique{J. Phys. (Paris)}%
\def\epjp{Eur.~Phys.~J.~Plus}%
\def\epjc{Eur.~Phys.~J.~C}%
\def\epl{Europhys.~Lett}%
\def\njp{New J.~Phys.}
\def\pdu{Phys.~Dark.~Univ.}
\let\astap=\aap
\let\apjlett=\apjl
\let\apjsupp=\apjs
\let\applopt=\ao

\preprint{ADP-25-32/T1294,TTP25-033}
\title{DAMA/LIBRA and dark matter: decisive tension or contrived cancellation}

\author{Giorgio Busoni}
\email{giorgio.busoni@adelaide.edu.au}
\affiliation{ARC Centre of Excellence for Dark Matter Particle Physics \& CSSM, Department of Physics, University of Adelaide, Adelaide, SA 5005}

\author{Jonathan M. Cornell}
\email{jonathancornell@weber.edu}
\affiliation{Department of Physics and Astronomy, Weber State University, 1415 Edvalson St., Dept. 2508, Ogden, UT 84408, USA}
\author{Will Handley}
\email{wh260@cam.ac.uk}
\affiliation{Cavendish Laboratory, University of Cambridge, JJ Thomson Avenue, Cambridge, CB3 0HE, UK}
\author{Felix Kahlhoefer}
\email{kahlhoefer@kit.edu}
\affiliation{Institute for Astroparticle Physics (IAP), Karlsruhe Institute of Technology (KIT), Hermann-von-Helmholtz-Platz 1, D-76344 Eggenstein-Leopoldshafen, Germany}
\author{Anders Kvellestad}
\email{anders.kvellestad@fys.uio.no}
\affiliation{Department of Physics, University of Oslo, N-0316 Oslo, Norway}
\author{Masen Pitts}
\affiliation{Department of Physics and Astronomy, Weber State University, 1415 Edvalson St., Dept. 2508, Ogden, UT 84408, USA}
\author{Lauren Street}
\affiliation{Department of Physics, University of Cincinnati, Cincinnati, OH 45221, USA}
\author{Aaron C. Vincent}
\email{aaron.vincent@queensu.ca}
\affiliation{
Department of Physics, Engineering Physics and Astronomy,\\
Queen’s University, Kingston ON K7L 3N6, Canada
}
\affiliation{%
Arthur B. McDonald Canadian Astroparticle Physics Research Institute, Kingston ON K7L 3N6, Canada
}
\affiliation{Perimeter Institute for Theoretical Physics, Waterloo ON N2L 2Y5, Canada}

\author{Martin White} 
\email{martin.white@adelaide.edu.au}
\affiliation{ARC Centre of Excellence for Dark Matter Particle Physics \& CSSM, Department of Physics, University of Adelaide, Adelaide, SA 5005}

\begin{abstract}
The ANAIS-112 and COSINE-100 experiments were constructed to test the long-observed dark matter-like annual modulation signal reported by DAMA. While they have reported null results in their annual modulation search, it remains possible that the combined effects of quenching, efficiency, resolution and binning could transform a common nuclear recoil rate into a signal that is visible in some detectors but not others. We assess the tension between DAMA/LIBRA and these latest experiments, under a range of hypotheses ranging from physical to general parameterisations of a common nuclear recoil input. We find that, in the most physically-motivated cases, the tension between DAMA and these other NaI experiments exceeds 5$\sigma$. Lowering the tension to reasonable values requires significant tuning, such as overfitting with large numbers of free parameters, and opposite-sign modulation between recoil signals on sodium versus iodine. 
 \end{abstract}

\maketitle

\section{Introduction}
It has been over two decades since the DAMA/LIBRA NaI experiment claimed the discovery of an annual modulation signal of dark matter (DM) at 6.3$\sigma$ \cite{Bernabei:2003za}. In combination with DAMA/LIBRA and DAMA/LIBRA phase2, 2.86 ton-yr of exposure over 22 annual cycles have now yielded a claimed signal over 13$\sigma$ \cite{Bernabei:2022ath}. Multiple analyses have shown that the energy dependence of this signal leads to a preference for a particle with a mass around 10 or 50 GeV, depending on whether it scatters primarily off sodium or iodine (e.g. \cite{Kelso:2013gda, Baum:2018ekm}). In parallel, however, advances in large direct detection experiments such as SuperCDMS, PICO-60, LZ and XENON1T have left very little room for a DM interpretation, with the current sensitivity being a full six orders of magnitude below DAMA/LIBRA's region of interest \cite{PhysRevLett.120.061802,PhysRevLett.121.051301,PhysRevLett.121.081307,PhysRevD.100.022001,PhysRevLett.131.041002,PhysRevLett.131.041003}. Nonetheless, the DAMA/LIBRA DM interpretation has persisted on the premise that no experiment to date has formally excluded it by searching for an annual modulation signature with the same technology, namely extremely pure NaI scintillator crystals, leaving the door open to DM particles with some inscrutable affinity towards sodium or iodine. While some exploratory efforts occurred starting from the early 2000s \cite{Ahmed:2003su,Kim:2012rza,DM-Ice:2016snk}, none of these experiments were able to completely exclude DAMA/LIBRA in a model-agnostic way. 

In recent years, a new generation of NaI experiments has been releasing data: the ANAIS-112 detector \cite{Amare:2019jul,Amare:2021yyu,Coarasa:2024xec,Amare:2025dfq}, located in Canfranc and COSINE-100 \cite{Adhikari:2018ljm,COSINE-100:2021xqn,COSINE-100:2019lgn,COSINE-100:2021zqh,Carlin:2024maf}, at Yangyang. Neither of these experiments alone has quite reached the 5$\sigma$ sensitivity to conclusively rule out the NaI-philic DM hypothesis of DAMA/LIBRA. A recent joint analysis \cite{COSINE-100:2025eyc} combining these datasets into two single-energy bins has excluded a modulation signal compatible with DAMA's at the level of 4.68$\sigma$ (3.53$\sigma$) in the 1-6 (2-6) keV range.  

\begin{figure*}[t]
  \includegraphics[width=\textwidth]{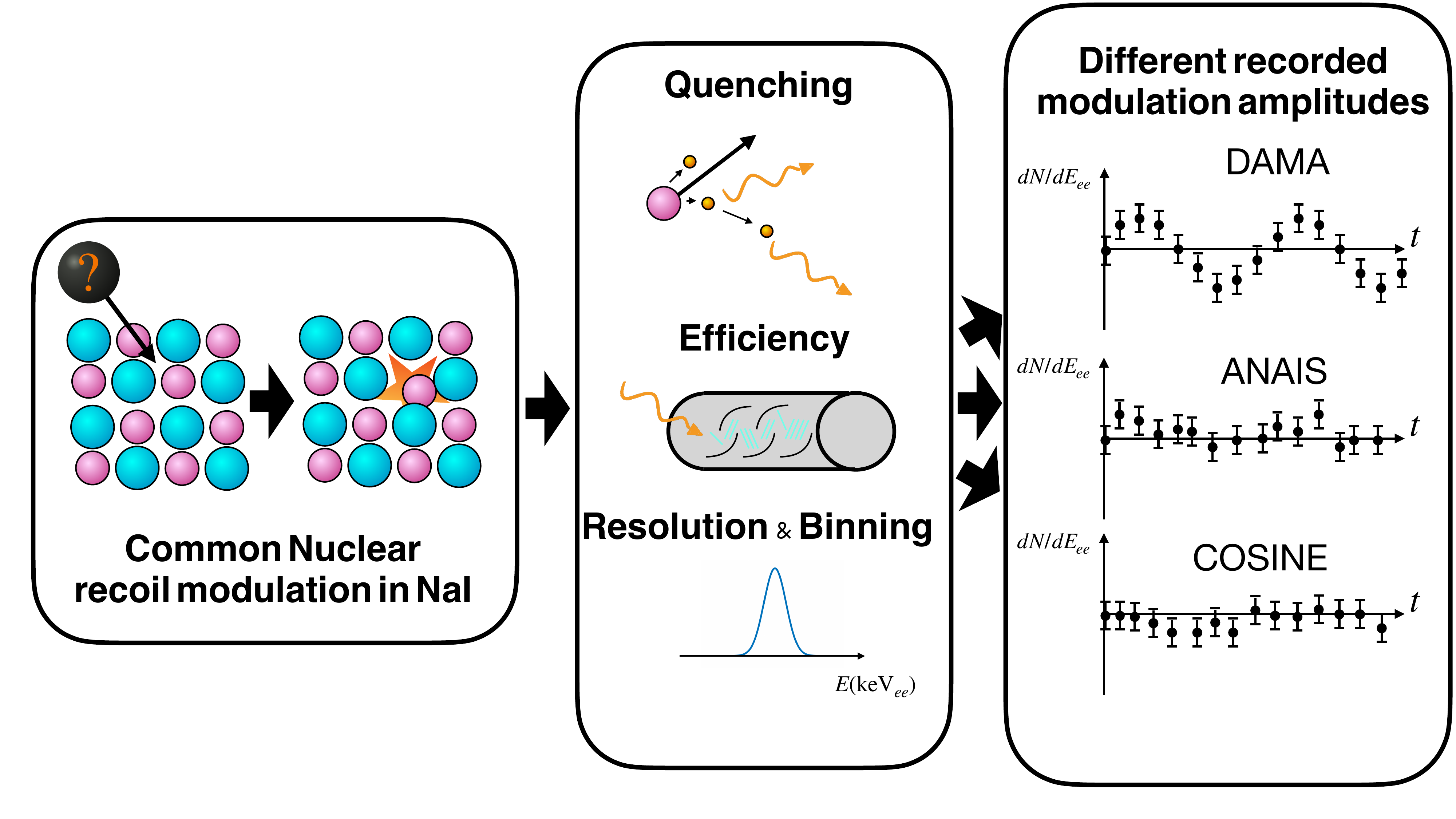}
  \caption{A schematic diagram of how we arrive at a predicted modulation spectrum. The calculation starts with a common nuclear recoil spectrum and takes into account differences in resolution, binning, and quenching factors to determine expected modulation rates, which could vary substantially between the various experiments.}
\end{figure*}

In addition, the two experiments have separately determined modulation amplitudes for events binned in energy \cite{Carlin:2024maf, Amare:2025dfq}, assuming the expected phase for DM scattering, a useful result for model discrimination. These can be directly compared to similar results previously reported by DAMA \cite{Bernabei:2021kdo}. Performing such a comparison will be the goal of this letter. We use the full spectral information reported by the different experiments and determine how well we can reconcile the spectra reported by DAMA/LIBRA and COSINE-100/ANAIS-112 if we assume the modulation is due to DM-nucleon scattering.  This comparison is not as trivial as it may seem: given a common \textit{nuclear recoil spectrum}, the differences in binning and energy resolution can lead to substantially different predictions for the \textit{measured} modulation  spectrum between the various experiments. To remain as general as possible, we adopt two different model-independent approaches: a DM effective field theory in which the nuclear recoil spectrum is calculable and a more general scenario where the nuclear recoil spectrum is parameterized with various functional forms. We undertake fits to the  experimental data sets to determine the best-fit parameters of these models, and based on these results we quantify the tension between the experiments. 

\title{Context}
The scattering rate of weakly-interacting DM particles from a nuclear target is given in terms of the differential cross section $d\sigma/dE_R$ by
\begin{equation}
\label{eq:recoil}
\frac{dR}{dE_R}=\frac{\rho_0}{m_\chi m_N}\int^{v_{\text{max}}}_{v_{\text{min}}}d^3v v \tilde{f}(\mathbf{v},t)\frac{d\sigma}{dE_R}
\end{equation}
where $\rho_0$ is the local DM mass density, $m_\chi$ is the DM particle mass, $m_N$ is the mass of the recoiling nucleus,  $v_{\max}$ is the escape velocity and $v_{\min}$ is the minimum velocity needed to cause a nucleus to recoil with energy $E_R$. $\tilde{f}(\mathbf{v},t)$ is the DM velocity distribution in the lab frame. The lab-frame distribution is related to the Galactic frame distribution $f(\mathbf{v})$ by a Galilean boost: $\tilde{f}(\mathbf{v})=f(\mathbf{v}+\mathbf{v}_{\text{obs}}(t))$,
where $\mathbf{v}_{\text{obs}}(t)=\mathbf{v}_{\odot}+\mathbf{V}_\oplus(t)$, $\mathbf{v}_{\odot}$ is the velocity of the Sun relative to the DM reference frame and $\mathbf{V}_{\oplus}$ is the velocity of the Earth about the Sun. 
In Boreal summer, the addition of the Earth's orbital velocity with that of the Sun boosts the ``wind'' of DM in our rest frame, whilst the wind is slower during the Austral summer. This changes the number of particles above the minimum speed required to create an observable nuclear recoil.
The resulting annual modulation~\cite{Drukier:1986tm,Spergel:1987kx,Freese:1987wu} is parameterised as
\begin{equation}
    \frac{dR}{dE_R}=A_0+A_1\cos\left[\omega(t-t_0)\right]+... 
    \label{eq:parametrisation}
\end{equation}
Experimental results are reported not in terms of the nuclear recoil energy $E_R$ but in terms of the measured electron-equivalent energy $E_{ee}$, which is related to the former by a quenching factor $Q_T(E_R)$ that depends on both the target $T$ and the recoil energy.
For a NaI detector with two potential nuclear targets $T=\{$Na,I$\}$, the total number of events in a given experimental bin $i$ is given by
\begin{align}
    N_i &= \int_{E_{i,\text{min}}}^{E_{i,\text{max}}} dE_{ee} \epsilon(E_{ee}) \times\\
    &\int_0^\infty dE_R \sum_T \xi_T\frac{1}{\sqrt{2\pi}\sigma(E_{ee})}e^{-\frac{(E_{ee}-Q_T(E_R)E_R)^2}{2\sigma(E_{ee})}} \frac{dR_T}{dE_R},\nonumber\label{eq:detectorresp}
\end{align}
where $\xi_T$ indicates the mass fraction of the target $T$, $dR_T/dE_R$ is the predicted recoil energy spectrum for that target, $E_{i,\text{min}},E_{i,\text{max}}$ define the bin energy range, $\epsilon$ is the detector efficiency and $\sigma$ is the detector resolution. 
As all collaborations report efficiency-corrected results, we will set $\epsilon=1$ for all detectors.

DAMA/LIBRA phases 1 and 2 \cite{Bernabei:2021kdo, Bernabei:2022xgg} have reported results compatible with a DM annual modulation signature in the energy range $1-6 \keVee$. However, they have presented results over a larger energy range, with DAMA/LIBRA-phase2 releasing data in the energy range $0.75-20 \keVee$ with a total exposure over 8 annual cycles of 1.53 t$\times$yr. This is in addition to data from DAMA/NaI and DAMA/LIBRA-phase1, which bring the total exposure in the range $2-20 \keVee$ to 2.86 t$\times$yr. In this work we fit to all of the available modulation data as presented in \cite{Bernabei:2021kdo}. The DAMA/LIBRA 
detector resolution is taken to be $    \sigma(E_{ee}) = a \sqrt{E_{ee}} + b E_{ee}$
with $a=0.488\sqrt{\keVee}$ and $b=0.0091$~\cite{DAMA:2008bis}.

COSINE-100 recently released results with $6.4 \mathrm{yr}$ of exposure \cite{Carlin:2024maf}, where they find no evidence of an annual modulation signal and a greater than $3\sigma$ tension with DAMA/LIBRA in the energy range $1-6 \keVee$. In our fits, we include their reported modulations in the range $0.75-20 \keVee$ and make use of the
detector resolution presented in \cite{Kang:2019uuj}:
$ \sigma(E_{ee}) = \sqrt{a E_{ee} + b E_{ee}^2}$
with $a = 0.081483 \keVee$, $b = 0.001885$.

ANAIS-112 recently presented binned modulation amplitude data from their 6-year exposure dataset \cite{Amare:2025dfq}. This data is incompatible with the
DAMA/LIBRA modulation signal in the $1-6 \keVee$ range at nearly $4 \sigma$ confidence level.
We fit to their reported modulation amplitudes over the range $1-20 \keVee$, and take the 
detector resolution to be $ \sigma(E_{ee}) = a + b \sqrt{E_{ee}} $
with $a = -0.08 \keVee$, $b = 0.378 \sqrt{\keVee}$~\cite{Amare:2018sxx}.

Equally important is the nuclear recoil quenching factor $Q_T(E_R)$, defined as the ratio of scintillation light yield produced by nuclear recoil to that of electron recoil at the same energy. We take recently-measured energy-dependent quenching factors for sodium from Ref.~\cite{Carlin:2024maf} and iodine from Ref.~\cite{Lee:2024unz}, for all experiments. We will, however, examine the impact of alternative quenching factors.

\section{Quantifying the tension between the datasets}
We follow the approach from \cite{Maltoni:2003cu} to quantify the tension between the datasets of different experiments.

Consider two experiments, $A$ and $B$, that measure a number of observables $m_A$ and $m_B$, respectively, and a model that describes the signal using $n$ parameters $c_i$.  
We define $\chi_X^2(c_i)$ as the $\chi^2$ obtained from the data of experiment $X=A,B$.  
We denote the best-fit point obtained by fitting the data of experiment $X$ alone as $c_i^X$, so that $\chi_X^2(c_i)$ has a minimum at $c_i=c_i^X$.  
Similarly, $c_i^{A+B}$ is the best-fit point obtained by fitting the data of both experiments simultaneously. In this case, the function  
\begin{equation}
   \chi_{A+B}^2(c_i) = \chi_A^2(c_i) + \chi_B^2(c_i)
\end{equation}
has a minimum at $c_i = c_i^{A+B}$, and the quantity  
\begin{equation}
   \delta\chi^2 = \chi_{A+B}^2(c_i^{A+B}) - \chi_A^2(c_i^{A}) - \chi_B^2(c_i^{B})
   \label{eq:tensionstat}
\end{equation}
follows a $\chi^2$ distribution with $n$ degrees of freedom.  

This can be used to determine the level of statistical inconsistency of the two datasets, under the assumption that the chosen model describes the signals with the right level of precision to reproduce the data of the experiments. We will use this statistical test to assess the compatibility between DAMA/LIBRA and the combination of COSINE-100 and ANAIS-112. We will employ two qualitatively different approaches to derive the tension between the DAMA/LIBRA and combined COSINE-100 and ANAIS-112 datasets. 

\section{Detailed particle astrophysics test} The first is to use concrete assumptions for the astrophysics and particle physics defined in Eq.~\eqref{eq:recoil}. The DM distribution $f(\mathbf{v})$, is well-described by a Maxwellian velocity distribution in the halo frame, with a peak velocity $v_{0} = 240\, \pm \,8$\,km\,s$^{-1}$ \cite{Reid:2014boa}, cut off at the escape velocity $v_{\rm{\text{esc}}} = 528 \pm 25$\,km\,s$^{-1}$, based on \emph{Gaia} data \cite{Deason:2019kgj}. The halo parameters are fixed to their central values, as varying these parameters is expected to affect all experiments in the same way. The potential impact of varying the velocity distribution will be implicitly captured by the more-model-independent parametrization discussed later.
On the particle physics side, we consider elastic scattering of DM on nuclei. However, instead of simply considering spin-independent and spin-dependent scattering, we adopt a more model-agnostic approach by considering an effective field theory (EFT) allowing for different types of interactions between DM and Standard Model particles. This approach avoids the need to specify the detailed microphysics of DM interactions, as long as the relevant energy scale is below the cutoff scale $\La$. For elastic scattering, we can simply set $\La$ equal to the hadronic scale, i.e.\ $\Lambda=2$~GeV without loss of generality. (A \textit{non}-relativistic EFT approach was taken with earlier data \cite{Kang:2019uuj,COSINE-100:2019wga, COSINE-100:2021xqn}. Relativistic operators were also fit to the earlier data in \cite{Kang:2019dbr}, but there only one operator was considered at a time, in contrast to our global fit.)

To construct the EFT, we further assume that the DM particle is a Dirac fermion and a singlet under the Standard Model gauge group. 
At the non-relativistic level, the interactions of all DM candidates map onto a finite set of Galilean-invariant operators. In this work we restrict ourselves to spin-1/2 dark matter for concreteness. While the interactions of scalar and vector dark matter arise from different relativistic operator bases, and hence lead to different correlations among the non-relativistic operator coefficients, the fermionic operators we consider already generate a broad set of relevant non-relativistic interaction structures. In particular, they include the dominant spin-independent and spin-dependent responses, as well as momentum- and velocity-suppressed interactions. We therefore expect that extending the analysis to spin-0 or spin-1 dark matter would not qualitatively change our conclusions, although it could shift the precise best-fit regions. Sec. \ref{sec:generic} covers more generic parametrisations of the signal.

Following the notation of Refs.~\cite{Bishara:2017nnn,Brod:2017bsw}, we write the interaction Lagrangian for the theory as
\begin{equation}
  \mathcal{L}_{\rm{int}} = \sum_{a,d} \dfrac{\C{a}{d}}{\La^{d-4}} \Q{a}{d}\,,
\end{equation}
where $\Q{a}{d}$ is a particular effective operator involving DM and Standard Model fields, $d\geq 5$ is the mass dimension of the operator and $\C{a}{d}$ is the
dimensionless Wilson coefficient associated to $\Q{a}{d}$. The full Lagrangian density for the theory is then
$ \mathcal{L} = \mathcal{L}_{\rm{SM}} + \mathcal{L}_{\rm{int}} + \overline{\chi}\left(i\slashed{\partial}-m_\chi\right)\chi\,,
$ such that the free parameters of the theory are the DM mass $m_\chi$, and
the set of dimensionless Wilson coefficients $\{ \C{a}{d} \}$. 

The phenomenology of DM in this model will generically be dominated by the lowest-dimension operators, unless their Wilson coefficients are accidentally suppressed or vanish, in which case higher-dimensional operators may become relevant. Under this assumption, we limit ourselves to operators with $d\le 6$. 
At dimension 5, there are the two dipole operators
\begin{align}
  \Q{1}{5} &= \frac{e}{8\pi^2} (\overline\chi \sigma_{\mu\nu} \chi) F^{\mu\nu} \,, \\
  \Q{2}{5} &= \frac{e}{8\pi^2} (\overline\chi i \sigma_{\mu\nu} \gamma_5 \chi) F^{\mu\nu} \, ,
\end{align}
where $F_{\mu\nu}$ is the electromagnetic field strength tensor and $e$ is the electromagnetic charge. These operators give rise to long-range interactions, i.e.\ steeply-falling recoil spectra.

At dimension six, we consider the operators
\begin{align} \label{dim6efts}
  \Q{1,q}{6} &= (\overline\chi \gamma_\mu \chi)(\overline{q} \gamma^\mu q)\,, \\
  \Q{2,q}{6} &= (\overline\chi \gamma_\mu \gamma_5 \chi)(\overline{q} \gamma^\mu q)\,, \\
  \Q{3,q}{6} &= (\overline\chi \gamma_\mu \chi)(\overline{q} \gamma^\mu \gamma_5 q)\,, \\
  \Q{4,q}{6} &= (\overline\chi \gamma_\mu \gamma_5 \chi)(\overline{q} \gamma^\mu \gamma_5 q)\,.
  \label{dim6efts:end}
\end{align}
The first two operators give rise to spin-independent interactions, while the last two give rise to spin-dependent interactions. Moreover, $\Q{1,q}{6}$ and $\Q{4,q}{6}$ are independent of the momentum transfer and the DM velocity, while $\Q{2,q}{6}$ and $\Q{3,q}{6}$ are suppressed in the low-velocity limit. Together these operators therefore capture a wide range of different possibilities for elastic scattering.

The effective operators are defined at the scale $\La=2$~GeV, where the Higgs, $W$ and $Z$ bosons as well as the top, bottom and charm quarks have been integrated out. We do not consider interactions with leptons, which do not give rise to nuclear scattering at tree-level. Following the assumption of Minimal Flavour Violation, we take the Wilson coefficients for operators involving the down and strange quarks to be equal, but we allow the Wilson coefficients for operators involving up quarks to differ. We parameterize the relative couplings between the $u$ and $d$-type quarks by angles $\theta^{(6)}_a$, in the following form:
\begin{align}
    C^{(6)}_{a,d} &= C^{(6)}_a \sin \theta^{(6)}_a = C^{(6)}_{a,s}\,, \\
                    C^{(6)}_{a,u} &= C^{(6)}_a \cos \theta^{(6)}_a \; .
\end{align}
In the EFT setup, we therefore have 11 model parameters to consider.  

The fits are performed with \gambit \cite{gambit, gambit_addendum} and its \darkbit~\cite{GAMBITDarkMatterWorkgroup:2017fax} module, using \diver 1.3\footnote{\url{https://github.com/diveropt/Diver/releases/tag/v1.3.0}}~\cite{ScannerBit} to explore the parameter space, \directdm~\cite{Bishara:2016hek, Bishara:2017pfq, Bishara:2017nnn} to match the effective operators introduced above onto the non-relativistic operators relevant for nuclear scattering, and \ddcalc\footnote{The rate calculations and likelihoods for the modulation experiments included in this paper will be part of the the upcoming \ddcalcthree public release.}~\cite{GAMBIT:2018eea} to evaluate nuclear form factors, calculate the differential event rate, and evaluate the experimental likelihoods.

The best-fit values for all parameters for each scan to the modulation spectrum from DAMA/LIBRA, both ANAIS-112 and COSINE-100, or all three experiments combined, are shown in Table \ref{tab:EFTparams}. The corresponding $\chi^2$ values and resulting tension is shown in Table \ref{tab:EFTchi2}. The combination of ANAIS-112 and COSINE-100 is found to rule out the DM nuclear recoil interpretation of the DAMA signal with a significance of $5.1\sigma$.

Our best fit to the DAMA data alone prefers interactions via the $C^{(6)}_{3}$ operator, which induces spin- and momentum-dependent interactions, with a DM mass of 184 GeV. We find an almost equally good fit at a mass of approximately 40 GeV. Fitting to ANAIS+COSINE, as well as to all three experiments, yields small couplings to all operators, i.e. no DM signal. 

Restricting the nuclear recoil energy range in DAMA below 7 keV$_{\mathrm ee}$ provides a best fit mass of 36 GeV. However, this restriction does not change the qualitative conclusions regarding the tension between DAMA and other experiments, which remains above 5$\sigma$, hence we choose to present the fit over the full range for completeness.

It has been speculated that quenching factors may be specific to individual crystals, rather than to the material itself.
Since DAMA has never measured the energy dependence of its crystal's quenching factor, they assume a constant value of $0.3$ for sodium and $0.09$ for iodine~\cite{Bernabei:1996vj}. Measurements from other groups have consistently shown a strong energy-dependence, with sodium quenching factors never being higher than $\sim 0.2$, and going as low as $0.1$ below 5 keV~\cite{Carlin:2024maf}. Nonetheless, we have repeated all scans using the DAMA quenching factors for the DAMA response modeling, and still find that this only modestly reduces the tension to 4.63$\sigma$. The final column of table \ref{tab:EFTchi2} shows the resulting $\chi^2$ values.  

\begin{figure*}[t!]
  \includegraphics[width=\textwidth]{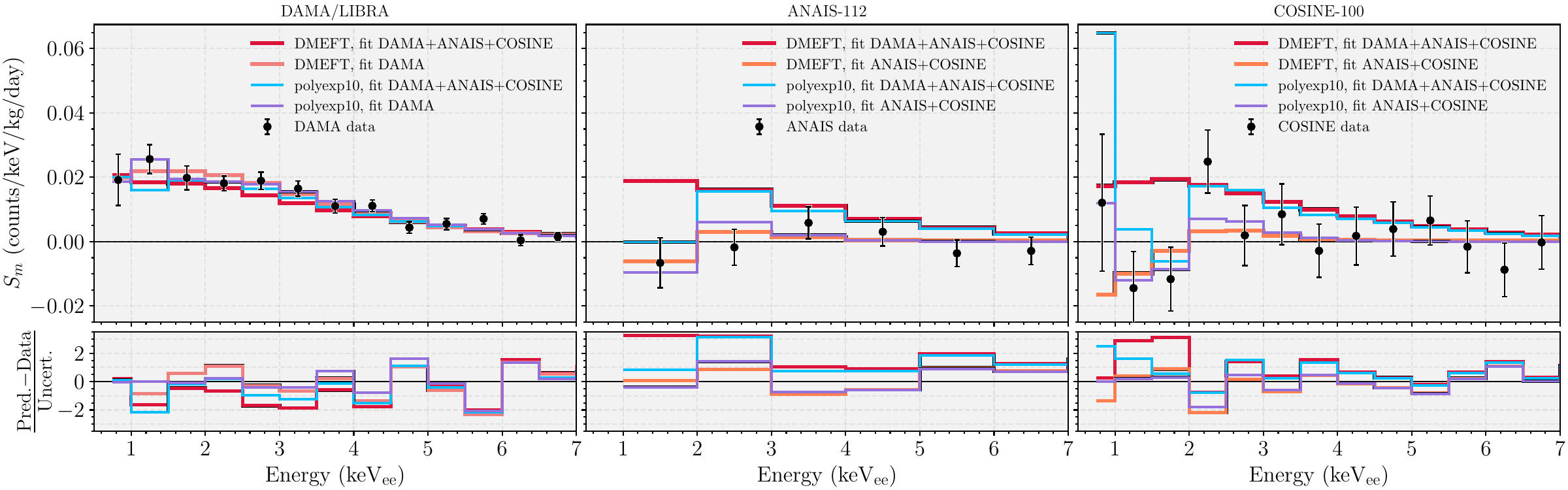}
  \caption{Comparison of experimental data to the predicted rates at the best-fit points of selected fits. \label{fig:bestfits}}
\end{figure*}

Fig.~\ref{fig:bestfits} shows the expected spectrum seen at DAMA/LIBRA (left), ANAIS-112 (centre) and COSINE-100 (right), if the signal is fit to that experiment only (orange), or to all three simultaneously (red). The large discrepancy between the data points (black) and the red curves for all three experiments illustrates the tension---in this EFT model, the strong preference of DAMA for modulation at low energies leads to similar predictions for the other experiments, a scenario which the data do not support.

In addition to the tension present in the modulation data, there is also tension in the predicted total rate. The best-fit point in the EFT setup predicts a total event rate in COSINE corresponding to around 8000 signal events in the bin [1, 1.5] keV, which should be compared to an observed number of 32294 events and a background expectation of $31572 \pm 1074$ events based on an exposure of 97.7 kg years~\cite{COSINE-100:2021xqn}. Hence, while the total rate predicted by the best-fit point cannot be excluded considering the total rate alone, it is in strong tension with data once the background model is taken into account. However, since these data points are not statistically independent from the modulation data, they cannot be easily combined with the tension presented above.

\begin{figure}
    \centering
    \includegraphics[width=1\linewidth]{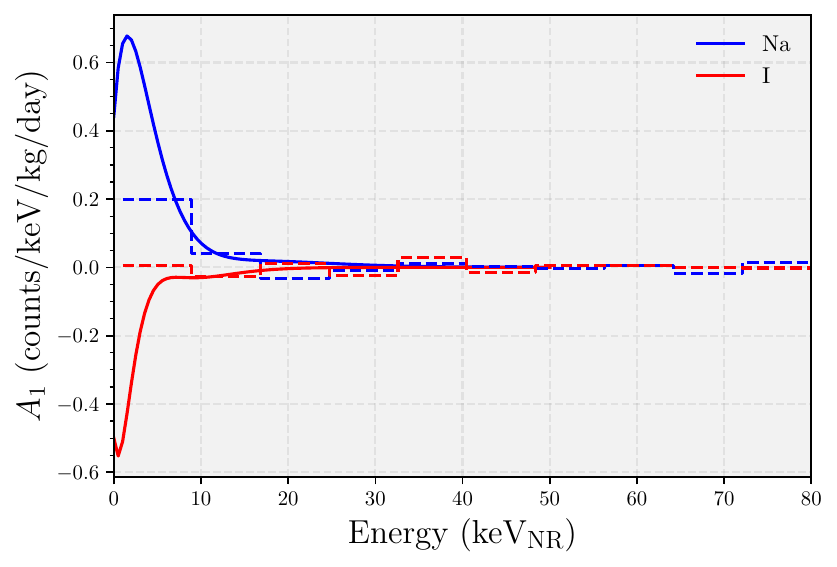}
    \caption{Best-fit nuclear recoil spectra necessary to produce ``only'' a 3$\sigma$ tension between DAMA/LIBRA and COSINE+ANAIS, using the polynomial-exponential parametrisation \eqref{eq:polyexp} (solid lines), and binned (2$\sigma$ tension, dashed lines) with 10 times 2 bins. The pathological opposite-sign recoil between Na and I is generic of models that improve the overall fit.}
    \label{fig:NRbestfit}
\end{figure}

\begin{table}
\caption{EFT Parameter values that maximize individual experimental likelihoods and the likelihood for all experiments combined. \label{tab:EFTparams}}
\begin{tabular}{lcccc}
\toprule
 & \bf All && \bf DAMA & \bf ANAIS+COSINE\\
\midrule
$m_\chi$ [GeV]            & 17.1                   && 184                   & 102                   \\
$C^{(5)}_1$               & $1.41 \times 10^{-4}$  && $1.47 \times 10^{-6}$ & $2.44 \times 10^{-5}$ \\
$C^{(5)}_2$               & $1.00 \times 10^{-8}$  && $1.04 \times 10^{-8}$ & $1.22 \times 10^{-8}$ \\
$C^{(6)}_{1}$             & $6.12 \times 10^{-5}$  && $9.08 \times 10^{-7}$ & $1.51 \times 10^{-6}$ \\
$C^{(6)}_{2}$             & $1.19 \times 10^{-7}$  && $8.18 \times 10^{-4}$ & $2.82 \times 10^{-8}$ \\
$C^{(6)}_{3}$             & $1.47 \times 10^{-7}$  && 0.436                 & $4.77 \times 10^{-7}$ \\
$C^{(6)}_{4}$             & $3.75 \times 10^{-7}$  && $1.13 \times 10^{-8}$ & $5.37 \times 10^{-8}$ \\
$\theta^{(6)}_1$          & $1.77\pi$              && $0.0353\pi$           & $0.676\pi$            \\
$\theta^{(6)}_2$          & $1.61\pi$              && $0.482\pi$            & $0.489\pi$            \\
$\theta^{(6)}_3$          & $1.89\pi$              && $1.30\pi$             & $1.27\pi$             \\
$\theta^{(6)}_4$          & $1.65\pi$              && $0.367\pi$            & $1.21\pi$             \\
\bottomrule
\end{tabular}
\end{table}

\begin{table}[h]
\caption{$\chi^2/\text{dof}$ values for the EFT fits to individual and combined experiments, as well as tension parameters. The first column shows the tension when all three experiments assume measured quenching factors for I and Na. The second column allows DAMA to use constant quenching factors.\label{tab:EFTchi2}}
\centering
 \begin{tabular}{l l l}
\toprule
 & \textbf{Common } & \textbf{Different quenching} \\
 & \textbf{quenching} & \textbf{for DAMA} \\
\midrule

All          & 181.36/86 & 173.37/86 \\
DAMA/LIBRA   & 74.02/28  & 71.83/28  \\
ANAIS+COSINE & 55.85/47  & 55.85/47  \\
\midrule
$\delta \chi^2/\text{dof}$ & 51.49/11 & 45.69/11 \\
p-value         & $3.37\times 10^{-7}$ & $3.66\times 10^{-6}$ \\
Tension         & $5.10 \sigma$ & $4.63 \sigma$ \\
\bottomrule
\end{tabular}
\end{table}

 \section{How generic is the tension?} \label{sec:generic} The use of specific particle physics and astrophysics assumptions means that any derived tension between DAMA/LIBRA and other experiments is necessarily model-dependent. We therefore also adopt a second approach which simply replaces the right-hand side of Equation~\ref{eq:recoil} by a generic functional form that is not constrained by any known particle, nuclear or astrophysics. We consider two sets of model-independent parameterisations: 1) A modulated nuclear recoil signal injected in $2N$ uniform $E_R$ bins between 1 and 80 keV\footnote{This range in nuclear recoil energy ensures that for quenching factors ranging from 5\% to 30\%, we cover the signal range in electron-equivalent energy.} allowing independent bin amplitudes between recoils with sodium and iodine, and 2) a polynomial $\times$ exponential parameterisation, inspired by the expected DM signal, but leaving coefficients completely free:
\begin{eqnarray}
            A_1 &=&   \sum_{T = \mathrm{Na,I}} \left(c^T_0 + c^T_1 E_R + c^T_2E_R^2 + ...\right)e^{-d_T E_R} ,
    \label{eq:polyexp}
\end{eqnarray}
where $d_T$ and the $c_i^T$ are free parameters. 
We use \diver 1.3 embedded in \gambitlight\footnote{\gambitlight is a lightweight version of \gambit, available at \url{github.com/GambitBSM/gambit_light_1.0}.} to maximise the likelihoods of the resulting spectra with respect to the DAMA/LIBRA data, combined ANAIS-112 and COSINE-100 datasets, and the combination of all three experiments,  and compute the tension statistic as in Eq.~\eqref{eq:tensionstat}. Demanding that the spectra be produced only by scattering with Na or I further worsens the tension in all cases. Resulting best fit $\chi^2$ values, p-values, and tension statistics are presented in the appendix. 

Using the tension statistic defined above, more than 12 bins (6 Na, 6 I) are required for the tension between DAMA/LIBRA and ANAIS-112 and COSINE-100 to fall below 5$\sigma$. With 20 bins, the tension reduces to 2$\sigma$. If we only allow for a signal in Na or I, the tension does not fall below 5$\sigma$ for 10 or fewer bins. Indeed, we find that the tension is only reduced when the contributions from Na and I in each bin have opposite 
signs (such that the one contribution peaks in summer, while the other peaks in winter), allowing for delicate cancellations to simultaneously fit the overall signal shape in all experiments thanks to their different binning, and therefore different mapping from nuclear to electronic recoil. 

The DM-inspired parameterisation from Eq.~\eqref{eq:polyexp} also leads to a reduced tension as the number of parameters is increased. When including only 4 free parameters (i.e. $c^{T}_0 e^{-d_T E_R}$ for both sodium and iodine), the tension is above the 5$\sigma$ level, but it drops modestly to 4.5, 4.1 and 3.2$\sigma$ as linear, quadratic and cubic terms in energy are added,  respectively. 

As in the binned case, the best-fit points to the combined data tend to be pathological, preferring a signal with a positive modulation amplitude for Na, and negative for I. In other words, while recoils off sodium peak in the summer, nuclear recoils on iodine would need to peak in winter, or vice-versa.  Fig.~\ref{fig:bestfits} shows the resulting signals in all three experiments in the case of the 10-parameter fit. The NR signal that leads to these results is in Fig.~\ref{fig:NRbestfit}. 

It is worth noting that this solution implies that the individual modulation amplitudes $A_1^T$ in both sodium and iodine must be much larger than the actually observed modulation amplitude $A_1 = \sum_T A_1^T$. However, the positivity of the time-dependent event rate in Eq.~\eqref{eq:parametrisation} implies that the average rate $A^T_0$ in each target must satisfy $A^T_0 \gtrsim |A^T_1|$. If $A_1^T$ has opposite sign for sodium and iodine, it follows that the total rate $A_0$ must be much larger than the observed modulation amplitude $|A_1|$. As discussed above, the total rate can be constrained using independent measurements from COSINE-100~\cite{COSINE-100:2021xqn}. This self-consistency requirement would place these scenarios under further strain.

\section{Conclusion}
We have revisited the tension between the DAMA/LIBRA, ANAIS-112 and COSINE-100 DM annual modulation datasets. For a Dirac fermion interacting with Standard Model particles through dimension-five and dimension-six operators, DAMA/LIBRA is in a 5.10 $\sigma$ tension with the other two experiments. This conclusion is now independent of the fact that  direct search DM experiments with other target nuclei would exclude the DAMA/LIBRA excess. If instead a generic parameterisation of the modulation amplitude vs nuclear recoil energy is used, the tension is still greater than 5$\sigma$ unless there is a significant degree of fine-tuning, resulting in a cancellation of the contributions from sodium and iodine. It remains interesting to consider a Southern hemisphere NaI experiment, such as the forthcoming SABRE South experiment~\cite{SABRE:2022twu}, which may shed further light on the mystery, in particular because it is expected to place a world-leading limit on the total rate of DM scattering in NaI detectors.

\begin{acknowledgements} We thank Ankit Beniwal, Torsten Bringmann, Joachim Brod, Jan Conrad, Andrew Fowlie, Hyun Su Lee, Seung Mok, Pat Scott, Sebastian Wild, Anthony Williams and Jure Zupan for useful input, along with the full GAMBIT collaboration. This work was performed using the Cambridge Service for Data Driven Discovery (CSD3), part of which is operated by the University of Cambridge Research Computing on behalf of the STFC DiRAC HPC Facility (www.dirac.ac.uk). The DiRAC component of CSD3 was funded by BEIS capital funding via STFC capital grants ST/P002307/1 and ST/R002452/1 and STFC operations grant ST/R00689X/1. DiRAC is part of the National e-Infrastructure. Additional support and computing resources from the Center for High Performance Computing at the University of Utah are gratefully acknowledged. GB is supported by the Australian Research Council
grant CE200100008. JMC acknowledges support from the Weber State University College of Science and Academic Resources and Computing Committee. In addition, he is grateful to the Mainz Institute of Theoretical Physics (MITP) of the Cluster of Excellence PRISMA$^+$ (Project ID 39083149), for its hospitality and partial support during the completion of this work. FK acknowledges support from the DFG via Emmy Noether Grant No.\ KA 4662/1--2. AK is supported by the Research Council of Norway (RCN) through the FRIPRO grant 323985 PLUMBIN'. MP was supported by the Jim S. Bateman Research Fund at Weber State University. ACV was supported by Arthur B. McDonald Canadian Astroparticle Physics Research Institute, NSERC the Canada Foundation for Innovation and the Province of Ontario. Research at Perimeter Institute is supported by the Government of Canada through the Department of Innovation, Science, and Economic Development, and by the Province of Ontario. MJW is supported by the Australian Research Council
grants CE200100008 and DP220100007.

\end{acknowledgements}

\bibliographystyle{JHEP_pat}
\bibliography{R2,modrefs}

\appendix

\section*{Appendix}

We present additional details on the results of our fits in Table~\ref{tab:bigtab}.

\onecolumngrid
\begin{table*}
\caption{$\chi^2/\text{dof}$ values obtained by maximizing individual experimental likelihoods and the likelihood for all experiments combined for the generic functional form recoil spectra, and the resulting p-values and tensions obtained with the $\delta \chi^2$ tension statistic in Eq.~\eqref{eq:tensionstat}.
\label{tab:bigtab}} 
\begin{tabular}{lc@{\hspace{5mm}}c@{\hspace{7mm}}cccc}
\toprule
Model & Pars. &  All & DAMA & COSINE+ANAIS & p-value & Tension ($\sigma$) \\
\midrule
\textbf{polyxexp} & 4 & 168.44/93 & 76.81/35 & 53.81/54 & $1.22 \times 10^{-7}$ & 5.16 \\
 (independent        & 6 & 162.43/91 & 73.67/33 & 52.81/52 & $2.82 \times 10^{-6}$ & 4.54 \\
 Na+I)        & 8 & 160.72/89 & 72.91/31 & 52.51/50 & $2.35 \times 10^{-5}$ & 4.07 \\
         & 10 & 156.06/87 & 73.08/29 & 52.52/48 & $7.19 \times 10^{-4}$ & 3.19 \\
         \hline
{Na only}       & 2 & 181.91/95 & 92.33/37 & 57.76/56 & $1.23 \times 10^{-7}$ & 5.16 \\
         & 3 & 168.10/94 & 76.90/36 & 57.60/55 & $2.41 \times 10^{-7}$ & 5.03 \\
         \hline
{I only}        & 2 & 188.31/95 & 86.25/37 & 57.83/56 & $2.49 \times 10^{-10}$ & 6.22 \\
         & 3 & 177.83/94 & 75.81/36 & 57.81/55 & $1.36 \times 10^{-9}$ & 5.95 \\
        \hline \hline
\textbf{bins}     & 2 & 204.77/95 & 103.65/37 & 58.02/56 & $4.36 \times 10^{-10}$ & 6.13 \\
(independent         & 4 & 194.57/93 & 75.94/35 & 56.54/54 & $1.05 \times 10^{-12}$ & 7.03 \\
    Na + I)     & 6 & 176.70/91 & 74.02/33 & 51.30/52 & $2.49 \times 10^{-9}$ & 5.85 \\
         & 10 & 159.90/87 & 59.66/29 & 45.60/48 & $3.69 \times 10^{-8}$ & 5.38 \\
         & 12 & 156.98/85 & 56.99/27 & 44.74/46 & $1.63 \times 10^{-7}$ & 5.11 \\
         & 16 & 139.75/81 & 52.38/23 & 39.54/42 &  $5.07 \times 10^{-5}$ & 3.89 \\
         &20 &	121.59/77 & 46.60/19 & 40.52/38 & $2.31 \times 10^{-2}$	&1.99 \\
         \hline
Na only & 1 & 236.01/96 & 143.34/38 & 58.60/57 & $5.32 \times 10^{-9}$ & 5.72 \\
         & 2 & 186.08/95 & 78.28/37 & 58.53/56 & $2.00 \times 10^{-11}$ & 6.60 \\
         & 5 & 178.33/92 & 74.71/34 & 51.92/53 & $6.22 \times 10^{-10}$ & 6.07 \\
         & 10 & 170.68/87 & 67.47/29 & 49.24/48 & $4.93 \times 10^{-8}$ & 5.33 \\
         \hline
I only   & 1 & 204.78/96 & 103.66/38 & 58.61/57 & $7.02 \times 10^{-11}$ & 6.42 \\
         & 2 & 195.03/95 & 76.64/37 & 58.27/56 & $8.81 \times 10^{-14}$ & 7.37 \\
         & 5 & 191.22/92 & 75.10/34 & 51.84/53 & $1.58 \times 10^{-12}$ & 6.97 \\
         & 10 & 172.06/87 & 72.06/29 & 47.69/48 & $9.97 \times 10^{-8}$ & 5.20 \\
\bottomrule
\end{tabular}
\end{table*}

\end{document}